\documentstyle[epsfig]{aprim}

\newif\ifAMStwofonts

\ifoldfss
  \ifCUPmtlplainloaded \else
    \NewTextAlphabet{textbfit} {cmbxti10} {}
    \NewTextAlphabet{textbfss} {cmssbx10} {}
    \NewMathAlphabet{mathbfit} {cmbxti10} {} 
    \NewMathAlphabet{mathbfss} {cmssbx10} {} 
  \fi
  \ifAMStwofonts
    \ifCUPmtlplainloaded \else
      \NewSymbolFont{upmath} {eurm10}
      \NewSymbolFont{AMSa} {msam10}
      \NewMathSymbol{\upi}     {0}{upmath}{19}
      \NewMathSymbol{\umu}     {0}{upmath}{16}
      \NewMathSymbol{\upartial}{0}{upmath}{40}
      \NewMathSymbol{\leqslant}{3}{AMSa}{36}
      \NewMathSymbol{\geqslant}{3}{AMSa}{3E}

    \fi
  \fi
\fi 

\ifnfssone
  \newmathalphabet{\mathit}
  \addtoversion{normal}{\mathit}{cmr}{m}{it}
  \addtoversion{bold}{\mathit}{cmr}{bx}{it}
  \newmathalphabet{\mathbfit} 
  \addtoversion{normal}{\mathbfit}{cmr}{bx}{it}
  \addtoversion{bold}{\mathbfit}{cmr}{bx}{it}
  \newmathalphabet{\mathbfss} 
  \addtoversion{normal}{\mathbfss}{cmss}{bx}{n}
  \addtoversion{bold}{\mathbfss}{cmss}{bx}{n}
  \ifAMStwofonts
    \ifCUPmtlplainloaded \else
      \UseAMStwoboldmath
      \makeatletter
      \new@mathgroup\upmath@group
      \define@mathgroup\mv@normal\upmath@group{eur}{m}{n}
      \define@mathgroup\mv@bold\upmath@group{eur}{b}{n}
      \edef\UPM{\hexnumber\upmath@group}
      \new@mathgroup\amsa@group
      \define@mathgroup\mv@normal\amsa@group{msa}{m}{n}
      \define@mathgroup\mv@bold\amsa@group{msa}{m}{n}
      \edef\AMSa{\hexnumber\amsa@group}
      \makeatother
      \mathchardef\upi="0\UPM19
      \mathchardef\umu="0\UPM16
      \mathchardef\upartial="0\UPM40
      \mathchardef\leqslant="3\AMSa36
      \mathchardef\geqslant="3\AMSa3E
    \fi
  \fi
\fi 

\ifnfsstwo
  \DeclareMathAlphabet{\mathbfit}{OT1}{cmr}{bx}{it}
  \SetMathAlphabet\mathbfit{bold}{OT1}{cmr}{bx}{it}
  \DeclareMathAlphabet{\mathbfss}{OT1}{cmss}{bx}{n}
  \SetMathAlphabet\mathbfss{bold}{OT1}{cmss}{bx}{n}
  \ifAMStwofonts
    \ifCUPmtlplainloaded \else
      \DeclareSymbolFont{UPM}{U}{eur}{m}{n}
      \SetSymbolFont{UPM}{bold}{U}{eur}{b}{n}
      \DeclareSymbolFont{AMSa}{U}{msa}{m}{n}
      \DeclareMathSymbol{\upi}{0}{UPM}{"19}
      \DeclareMathSymbol{\umu}{0}{UPM}{"16}
      \DeclareMathSymbol{\upartial}{0}{UPM}{"40}
      \DeclareMathSymbol{\leqslant}{3}{AMSa}{"36}
      \DeclareMathSymbol{\geqslant}{3}{AMSa}{"3E}
    \fi
  \fi
\fi 

\ifCUPmtlplainloaded \else
  \ifAMStwofonts \else 
    \def\upi{\pi}
    \def\umu{\mu}
    \def\upartial{\partial}
  \fi
\fi

\title[Manuscript Template]{The Large Adaptive Reflector concept}
\author[Dougherty et al.]
       {S.M. Dougherty$^1$, P.E. Dewdney$^1$, A. Gray$^1$ and A.R. Taylor$^2$\\
        $^1$National Research Council, Herzberg Institute for
        Astrophysics, Dominion Radio Astrophysical Observatory,
        Canada\\ $^2$Department of Physics \& Astronomy, University of
        Calgary, Canada} \date{}

\pagerange{\pageref{firstpage}--\pageref{lastpage}}
\pubyear{2005}

\begin{document}

\maketitle

\label{firstpage}

\begin{abstract}
Cost effective, new antenna technology is required to build the large
collecting area being planned for the next generation of radio
telescopes. The Large Adaptive Reflector (LAR) is a novel concept
being developed in Canada to meet this challenge.  A prototype with a
200-350m diameter reflector, operating up to 2 GHz with a bandwidth of
750 MHz is planned. With a collecting area up to $\sim10$\% of the
planned SKA, and an array feed capable of imaging a 0.3~deg$^2$
field-of-view, the prototype would address a number of the most
compelling questions in modern astrophysics.
\end{abstract}

\section{The LAR}
The LAR is a concept for a large diameter, large f/D parabolic
reflector that requires an air-borne platform to support the prime
focus receiver system. The active reflector surface has a very flat
profile, supported over a large area, to reduce significantly the cost
per unit area compared to traditional designs. A large field-of-view
across a wide bandwidth is attained through the use of a focal plane
phased array (Dewdney et al. 2002; Carlson et al. 2000).
\begin{figure}
 \begin{center}
   \epsfig{file=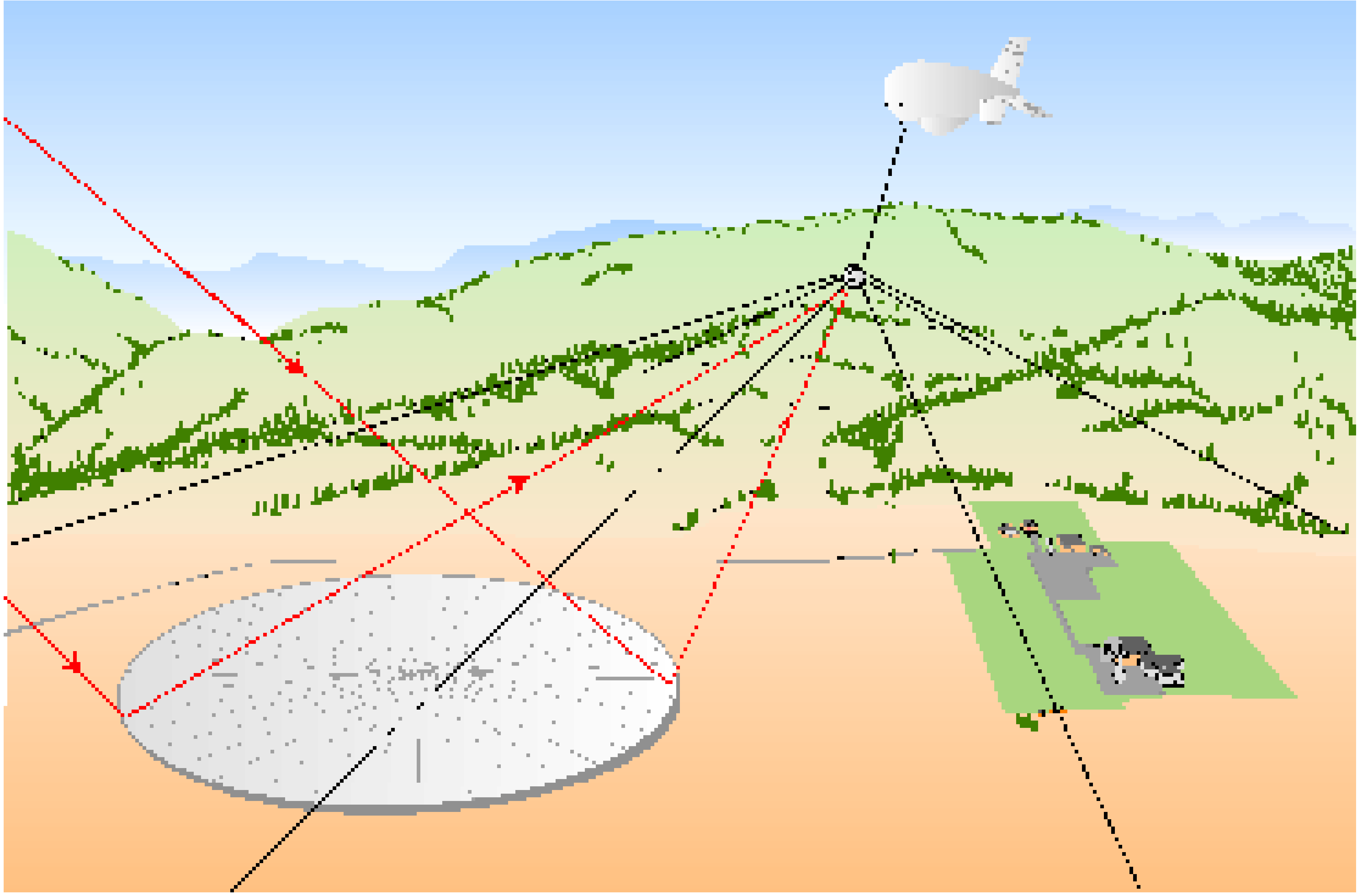,width=8.0cm}
   \epsfig{file=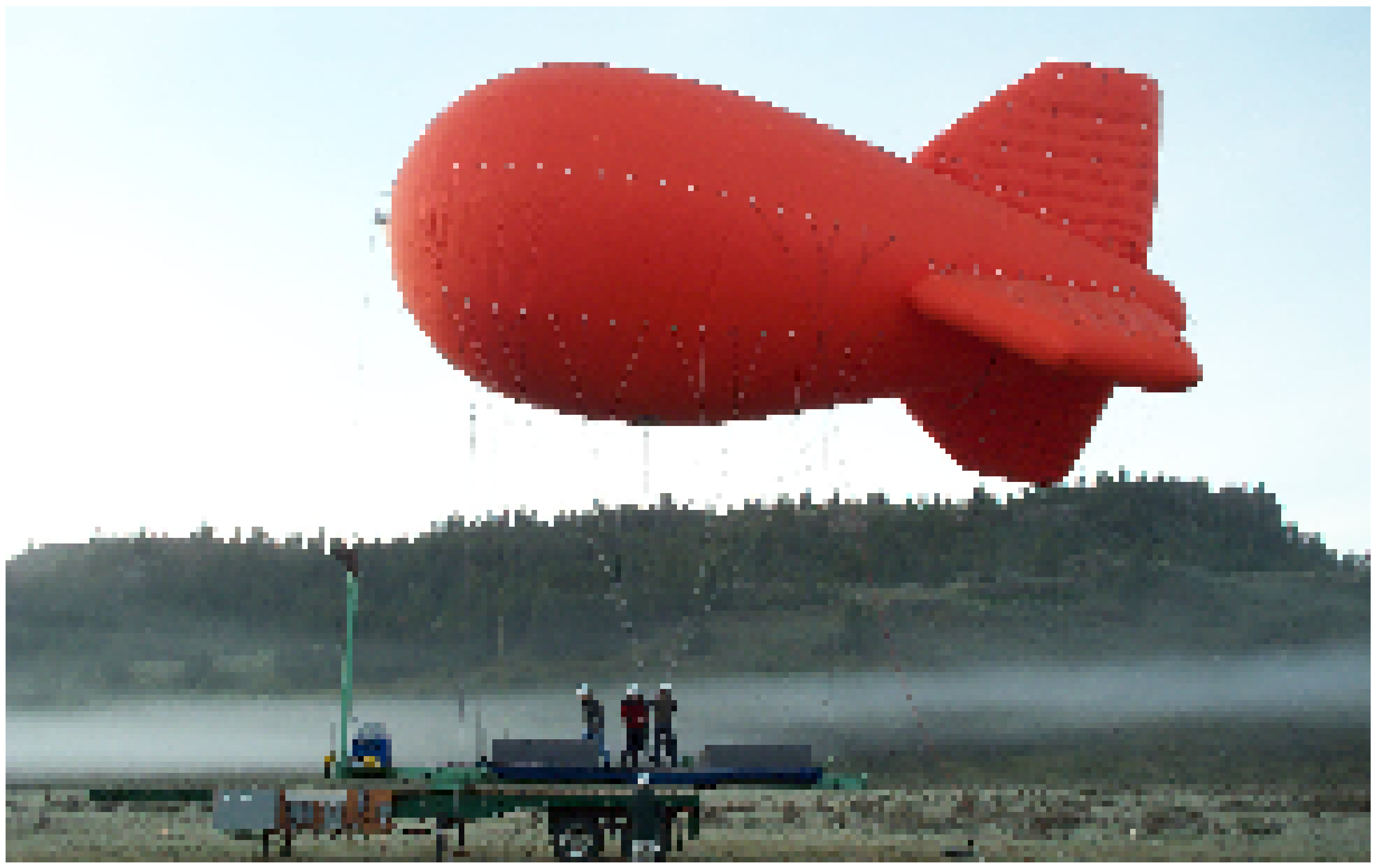, width=8.0cm}
 \end{center}
 \caption{The Large Adaptive Reflector Concept.  The top figure shows
 a cartoon of the system configuration. The lower figure shows the
 aerostat being used currently in the 1/3-scale prototype of the
 tethered aerostat system. \label{concept}}
\end{figure}
\subsection{Tethered Aerostat System}
As the LAR reflector has a focal length of several hundred metres,
the Focal Point Apparatus (FPA) needs to be supported by a
``sky-crane'', for which we have chosen a helium-filled aerostat
(Fig.~\ref{concept}). The aerostat is attached to the FPA via a leash,
isolating the FPA from balloon motions, and the position of the FPA is
controlled by six computer-controlled winches. This system is being
evaluated using a 1/3-scale prototype (Fig.~\ref{concept}) to verify a
computer simulation of the system. Once verified, the computer model
becomes an extremely powerful design tool for the full-scale system.
Recent tests on the prototype reveal that this system can work down to
a zenith angle of $60^\circ$, while maintaining the position of the
FPA to within a few centimetres, adequate for the focus design we
propose (Sec.~\ref{FPPA}).
\subsection{Reflector}
\begin{figure}
 \begin{center}
   \epsfig{file=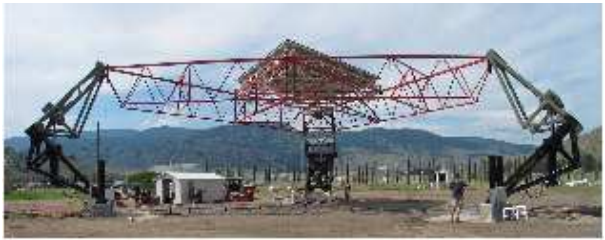,width=8.0cm}
 \end{center}
 \caption{A full-scale prototype LAR reflector facet\label{reflector}.}
\end{figure}

The LAR reflector is a large diameter, faceted approximation to an
offset parabolic reflector with a long focal length
(f/D$\sim$2.5). Each of the 20-m triangular facets is supported on
actuated legs that extend up to 12m. These actuators maintain the
parabolic shape of the reflector as the pointing direction is
changed. The major advantage of such a design is that the reflector is
very flat, permitting its weight to be supported at many locations,
rather than at the single or double mount points of more traditional
designs.  A single facet using this design has been built and tested
(Fig~\ref{reflector}). This has demonstrated the low cost of the
actuation design, and established an initial reflector cost estimate
of $\sim\$400$~US per square metre.
\subsection{Focal-Plane Phased Array}
\label{FPPA}
A phased-array feed at the focus confers a larger field-of-view (FOV)
than a traditional single-feed receiver system, by sampling the
electromagnetic field at the focus to form multiple beams.  The
ability to modify the beam pattern of an array feed is essential to
the LAR concept to control the illumination of the reflector as the
zenith angle changes.

Until recently, array feeds in radio astronomy have been impractical
because suitable array elements were either not available or were too
expensive. Today, the Vivaldi antenna is recognised as having the most
suitable characteristics, with wide bandwidth, packing density, and
low loss performance.

\begin{figure}
 \begin{center}
   \epsfig{file=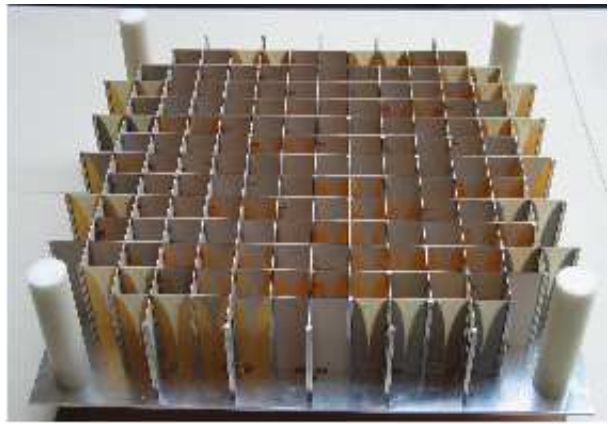,width=8.0cm}
 \end{center}
 \caption{A Vivaldi array \label{PHAD}.}
\end{figure}

A three-fold development process for attaining a phased-array feed
design for the LAR is planned: 1) Build a 200-element Vivaldi array
equipped with a simple, low cost, narrow band receiver system from
off-the-shelf components (Fig.~\ref{PHAD}). This is an engineering
demonstrator to establish a fundamental understanding of the
capabilities and limitations of phased-array feeds on reflector
telescopes; 2) In parallel, develop high performance uncooled,
monolithic, integrated LNAs to attain good system noise without the
need for cryogenically cooled amplifiers, and RF-to-Optical modules
for data transmission from the receivers to the ground-based
beamformer; 3) Lastly, develop an astronomically capable phased-array
as a prototype for the LAR.  Currently, the first two phases of this
development strategy are underway.

\subsection{The Focal Point Apparatus}
The Focal Point Apparatus (FPA) is the mechanism at the focus of the
LAR reflector that is supported by the aerostat, and controls the
pointing of the phased-array feed. The tethers that control the
position of the focus are attached to the FPA.  Taking into account
the necessary degrees of freedom, the workspace volume and a minimal
mass target, a number of well-known control mechanisms are ruled out,
and point to a cable-based mechanism acting within a large space
frame. A prototype of this system is currently being built, to be flown
on the tethered aerostat system in June 2006.
\section{Science with a CLAR}
A prototype of the LAR concept, known as the CLAR, is envisaged to be
a 300-m diameter reflector, with sky coverage down to $30^\circ$
elevation, an array feed with a FOV of 0.3~deg$^2$, and operating up
to 2 GHz with an instantaneous bandwidth ratio of 2:1. Such a prototype
represents $\sim7$ times the collecting area of the VLA, with 100
times its survey speed. With $\sim7$\% of the area of the SKA and a large
FOV giving high survey speed, this prototype can address many of the
compelling questions facing modern astrophysics.

In particular, the CLAR will measure the redshift and distribution of
millions of galaxies out to a redshift of z$\sim2$ using the 21cm
hyperfine transition of atomic hydrogen. This will enable the
measurement of the spatial power spectrum of the distribution of
galaxies at different values of z, opening up the way to determine
the equation of state of the universe. A more shallow survey covering
the whole sky will map the large-scale structure of the nearby
universe through the detection of over 10 million galaxies, surpassing
the SLOAN survey by at least a factor of 20.

With a collecting area of nine times the GBT and sky coverage down to
$30^\circ$ elevation, the CLAR will be a second-to-none pulsar
discovery machine. This will open up a new era in precision tests of
General Relativity through the binary pulsar systems that will
undoubtedly be found, as well as studies of the extreme physics of
neutron stars themselves.


\label{lastpage}

\clearpage


\begin{thebibliography}{99}
	\bibitem[\protect\citename{Dewdney et al. }2002]{Dewdney02} 
Dewdney P.E., Nahon M., Veidt B., 2002, Can. Aero. Space J., 28, 239

	\bibitem[\protect\citename{Carlson et al. }2000]{Carlson00} 
Carlson B., Bauwens K., Belostotski L., Cannon W., Chang Y.-Y.,  Deng X.,
Dewdney, P.E., Fitzimmons, J., Halliday D., Kurshner K., Lachapelle G., Lo D.,
Mousavi P., Nahon M., Shafai L., Steimer S., Taylor A.R., Veidt B., 2000,
Proc. SPIE Int. Symp on Astronomical Telescopes and Instrumentation, p 4015.

\end{thebibliography}
\end{document}